# Is Covid-19 severity associated with ACE2 degradation?

Ugo Bastolla[1*], Patrick Chambers[2], David Abia[1], Maria-Laura García-Bermejo[3] and Manuel Fresno[1]

[1]Centro de Biología Molecular Severo-Ochoa, CSIC-UAM, Madrid Spain

[2]Department of Pathology, Torrance Memorial Medical Center, USA

[3]Instituto Ramón y Cajal de Investigaciones sanitarias, Madrid, Spain

[*]Correspondence to ubastolla@cbm.csic.es

Covid-19 is particularly mild with children, and its severity escalates with age. Several theories have been proposed to explain these facts. In particular, it was proposed that the lower expression of the viral receptor ACE2 in children protects them from severe Covid. However, other works suggested an inverse relationship between ACE2 expression and disease severity. Here we try to reconcile seemingly contradicting observations noting that ACE2 is not monotonically related with age but it reaches a maximum at a young age that depends on the cell type and then decreases. This pattern is consistent with most existing data from humans and rodents and it is expected to be more marked for ACE2 cell protein than for mRNA because of the increase with age of the protease TACE/ADAM17 that sheds ACE2 from the cell membrane to the serum.

The negative relation between ACE2 level and Covid-19 severity at old age is not paradoxical but it is consistent with a mathematical model of virus propagation that predicts that higher viral receptor does not necessarily favour virus propagation and it can even slow it down. More importantly, ACE2 is known to protect organs from chronic and acute inflammation, which are worsened by low ACE2 levels. Here we propose that ACE2 contributes essentially to reverse the inflammatory process by downregulating the pro-inflammatory peptides of the angiotensin and bradykinin system, and that failure to revert the inflammation triggered by SARS-COV-2 may underlie both severe CoViD-19 infection and its many post-infection manifestations, including the multi-inflammatory syndrome of children (MIS-C). Within this view, lower severity in children despite lower ACE2 expression may be consistent with their higher expression of the alternative angiotensin II receptor ATR2 and in general of the anti-inflammatory arm of the Renin-Angiotensin System (RAS) at young age.

**Introduction**

The Covid-19 pandemics (Zhou et al. 2020) has overcome the startling figure of two million confirmed deaths (Dong et al. 2020), and excess death indicates that the real number is substantially higher (Vestergaard et al. 2020; Weinberger et al. 2020). However, in contrast with the high mortality of the elderlies, children predominantly develop a mild form of Covid-19 and their mortality is very low, as part of a general trend of increasing Covid-19 severity with age. Age-stratified mortality rates with respect to detected cases increase very steeply with age, and even more steeply if we consider that most infections are asymptomatic at young age and they tend to go undetected. Fig.1 shows the approximately exponential increase with age of detected cases, hospitalizations, ICU and deaths with respect to the number of seropositives detected in population-wide antibody surveys in Spain at 10 of May 2020. One can see that the increase with age is apparent at all levels of severity, except for a drop of ICU at high age that unfortunately is probably explained by the saturation of the health system during the first pandemic wave.

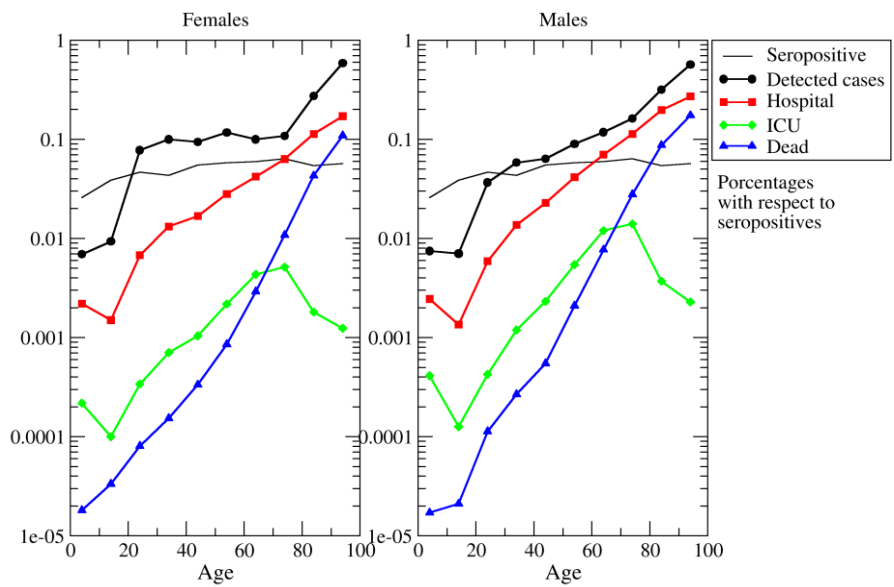

Fig.1 Covid-19 severity increases exponentially with age. Black: Cases detected through positive PCR. Red: PCR-positive patients that required hospitalization. Green: PCR-positive patients in ICU. The drop at high age probably reflects the saturation of the health system during the first wave of the pandemics. Black: Deceased patients with positive PCR (the real toll estimated with excess deaths is unfortunately much higher).

Several theories have been proposed to explain why Covid-19 is so mild with children, recently reviewed by Zimmerman and Curtis (2020). One group of theories postulates different immune response in children. Cristiani et al. (2020) invoke the strength of the innate immunity of children, which is further reinforced by the frequent infections and the vaccines that they are exposed to, which enhance their trained immunity. Carsetti et al. (2020) propose that the immune system of children responds better to new pathogens. Also at the level of the adaptive immunity children might have an advantage, since they present more often antibodies against common human coronaviruses that can cross-react with SARS-COV-2 (Ng et al. 2020). However, despite the immune response of children and adults to SARS-CoV-2 are markedly different, with children expressing a different spectrum of immune molecules, Pierce et al.

(2020) argued that the greater severity of hospitalized adults compared to children could not be attributed to their less efficient immune responses.

A second group of theories is based on factors that put aged adults at increased risk, including co-morbidities and, above all, differences in the endothelial system, as reported in a Nature news article (Cyranoski 2020). The uncompromised state of the endothelial system of children may protect them from the most severe complications of Covid-19 that originate from endothelial inflammation and dysfunction, as proposed by Monagle group at Melbourne hospital and by other groups (Nature News 2020).

However, the theory that captured most attention is probably the one based on the variation of the viral receptor ACE2 with age. It was recently observed that ACE2 mRNA and serum protein is lower in children than in adults (Bunyavanich, Do and Vicencio 2020; Muus et al. 2020; Sharif-Askari et al. 2020, Pavel et al. 2020), and it was proposed that lower receptor levels protect children from severe SARS-CoV-2 infection. Nevertheless, data on ACE2 expression across age are contradictory. Some works found that ACE2 protein in cells is lower in children than in adults (Inde et al. 2020) and others reached the opposite conclusion (Zhang, Guo et al. 2020; Ortiz-Bezara et al. 2020).

Here we go beyond the dichotomy between children and adults and propose an alternative interpretation of the data, supported by observations that indicate that ACE2 mRNA and protein expression start from zero during foetal life and reach a maximum at young age (Muus et al. 2020; Inde et al. 2020), after which they decay with age in adult rats and mice (Xie et al. 2006; Schouten et al. 2016; Yoon et al. 2016; Booeshaghi and Pachter 2020) and in humans (Chen et al. 2020; Zhang, Guo et al. 2020), with strong inter-individual variation and cell-type dependent maximum age (Ortiz-Bezara et al. 2020; Inde et al. 2020). ACE2 cell protein level behaves differently from ACE2 mRNA, due to the shedding of ACE2 from the cell membrane to the serum produced by the metalloprotease TACE/ADAM17 (Lambert et al. 2005). The increased expression of ADAM17 through age (Dou et al. 2017; Liu et al. 2019) predicts increasing ACE2 shedding and implies that the maximum expression is achieved at earlier age for ACE2 protein than for ACE2 mRNA.

Whereas the comparison between children and adults supports a positive correlation between ACE2 level and disease severity, the comparison between young and old adults supports their negative correlation. That lower receptor level is not necessarily a protective factor is predicted by a mathematical model according to which, in some circumstances, viruses propagate in the infected organism more slowly with higher receptor level (Ortega-Cejas et al. 2004; see Fig.4 for illustration). Most importantly, low ACE2 levels expose the lungs to acute inflammation (Imai et al. 2005), andACE2 is low in most common chronic pathologies including hypertension, angiocardiopathy, type 2 diabetes, chronic renal failure, pulmonary diseases and liver diseases (Li et al. 2020; Pagliaro and Penna 2020). For these reasons, several authors proposed that the severity of Covid-19 is exacerbated by the degradation of ACE2 by the virus (Sun et al. 2020; Gurwitz 2020; Verdecchia et al. 2020; Ciaglia et al. 2020; Offringa et al. 2020;

Annweiler et al. 2020 among others). In this paper, we propose that one of the main functions of ACE2 consists in contributing to reverse the inflammatory state, and that the degradation of ACE2 by SARS-CoV-2 is one of the main triggers of the acute inflammation observed in severe Covid-19.

**ACE2 in the context of the angiotensin and bradykinin systems**

The Angiotensin converting enzyme 2 (ACE2), the receptor of SARS, SARS-CoV-2 and other coronaviruses, plays a regulatory role in the RAS and in the Kallikrenin-kinin system (KKS). These two systems are strongly coupled, since they share some of their main functions (control of blood pressure, blood coagulation, control of inflammatory processes, regulation of the immune system response after infection or traumatic events) and two key enzymes, the carboxy-peptidases ACE and ACE2 that regulate the signal peptides of both systems, the family of angiotensin for the RAS and bradykinin for the KKS. Their receptors, which belong to the big family of G-protein coupled receptors (GPCR), may act in a synergistic manner, forming complexes that mutually enhance each other signalling through allosteric interactions (Quitterer and AbdAlla 2014). Thus, it can be considered that RAS and KKS form an integrated system.

In the RAS, ACE2 downregulates the pro-inflammatory and vasoconstrictor peptide Angiotensin II (Ang-II, which we denote here as Ang1-8 to make more explicit that it consists of the first 8 amino acids of the peptide Angiotensin I or Ang1-10) that increases blood pressure, transforming it into the form Ang1-7 that mediates vasodilator and anti-inflammatory effects through the Mas receptor. Moreover, ACE2 transforms Ang1-10, the precursor of Ang1-8, into the form Ang1-9 that is later transformed into Ang1-7 by ACE. In the KKS, ACE2 degrades Des-Arg-Bradykinin and Lys-Des-Arg-Bradykinin (BK1-8 and LysBK0-8), two peptides that signal through the receptor BKR1 that is expressed upon inflammation. Bradykinin signalling produces vasodilation, decreases blood pressure, increases vascular permeability and induces capillary leakage. These actions can produce the edemas suffered by severe Covid-19 patients. Moreover, bradykinin enhances the inflammation by recruiting neutrophils and leukocytes and by upregulating the protease TACE/ADAM17 that activates the tumor necrosis factor $\alpha$ (TNF-$\alpha$) and removes ACE2 from the cell membrane. The involvement of the KKS in Covid-19 cases has been discussed in recent publications (Nicolau, Magalhães and Vale 2020; van de Veerdonk et al. 2020; Garvin et al. 2020; Zwaveling et al. 2020). It is consistent with the observation that a fraction of Covid-19 patients present hypotension (66% of critical patients, Michard and Vieillard-Baron 2020, and 8% of hospitalized patients compared with 39% with hypertension, Table 2 of Lala et al. 2020).

We represent in Fig.2 the reduced molecular network of the RAS and KKS.

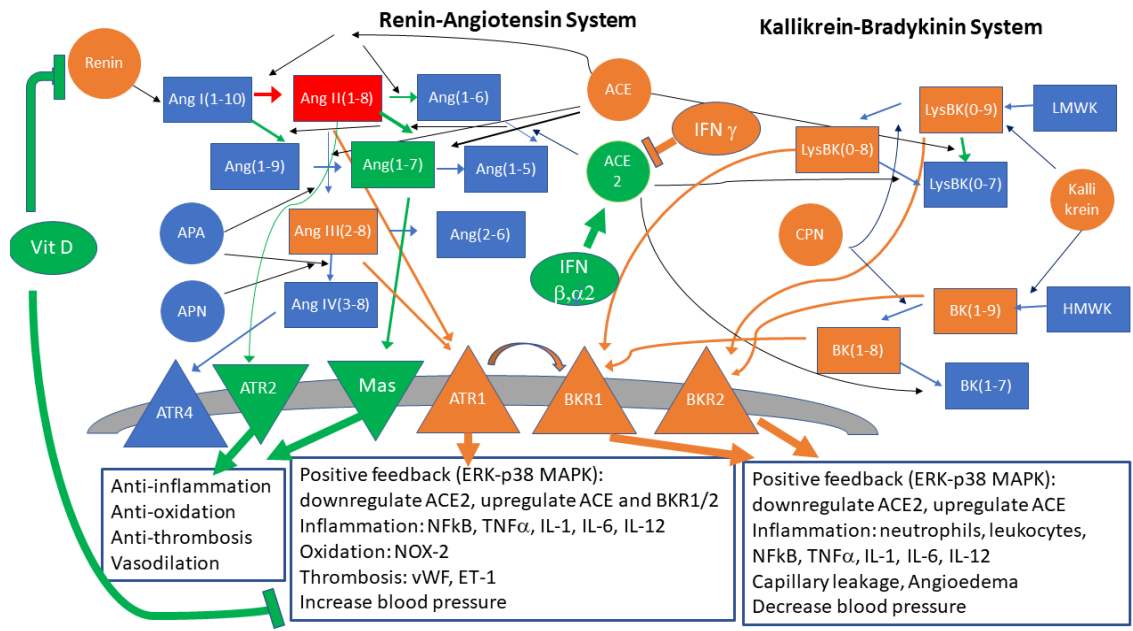

Fig.2 Simplified representation of the Renin-Angiotensin system and the Kallikrein-Bradykinin system, coupled through the peptidases ACE and ACE2 that act on both families of peptides and through the allosteric interactions between the G-protein coupled receptors of both families. Peptides of the angiotensin (Ang) and bradykinin (BK) family are represented as rectangles, coloured arrows indicate their transformations and circles indicate the peptidases that catalyse them (black arrows). Membrane receptors are indicated as triangles and the boxes indicate the main effects of their signalling. The colour code represents inflammatory effects (orange: pro-inflammatory; green: anti-inflammatory, blue: Neutral or not clear at present). Also represented as ellipses are Vitamin D, which downregulates Renin and the down-stream effectors NFkB and NOX, and Interferons $\alpha$ and $\beta$, both of which upregulate ACE2 with anti-inflammatory effects, and $\gamma$, which downregulates ACE2.

**The main functional role of ACE2 may consist in reversing the inflammation process**

ACE2 counteracts high blood pressure in the RAS and low blood pressure in the KKS, but it has anti-inflammatory effects in both systems. Interestingly, several studies have shown that ACE2 knockout mice do not present significant hypertension or cardiac anomalies (Crackower et al. 2002; Gurley et al. 2006; Alghamri et al. 2013), but they show enhanced response to Ang1-8 stimulation (Alghamri et al. 2013) and they develop lung edema after acute-inflammation (Imai et al. 2005). Edema is not the mechanical consequence of increased blood pressure, which does not happen in ACE2 KO mice, but it may derive from increased vascular permeability due to high bradykinin levels (van de Veerdonk et al. 2020). This is supported by the observation that attenuation of pulmonary ACE2 activity impairs inactivation of the DABK/BK1R axis and facilitates neutrophil infiltration (Sodhi et al. 2017). These observations connect ACE2 with the termination of the inflammatory process, as further discussed below.

A crucial link between ACE2 and the inflammatory process goes through the Tumor necrosis factor alpha (TNF-$\alpha$) converting enzyme (TACE), also known as ADAM17. This metallo-protease sheds ACE2 from the cellular membrane to the serum (Lambert et al. 2005), from which it is rapidly eliminated through urine, and at the same time it activates the cytokine TNF-$\alpha$, initiating a cascade process that, through the transcription factor nuclear factor $\kappa$ B (NF-$\kappa$B) promotes the activation of the inflammatory response. Importantly, ADAM17 is upregulated by the Ang1-8 receptor ATR1 (Deshotels et al. 2014; Xu et al. 2017) and the bradykinin receptor BKR1 (Parekh et al. 2020), both of which are downregulated by ACE2, which establishes two dangerous positive loops that can enhance ACE2 degradation.

Here we summarize the main steps with which the RAS and the KKS participate in the inflammation process

1) The infection activates the RAS, enhancing Ang1-8 production

2) High Ang1-8/ATR1 generates a positive feedback by upregulating ACE, downregulating ACE2 (Koka et al. 2008), and upregulating AT1R/ADAM17 (Deshotels et al. 2014), which sheds ACE2 from the cell membrane to the serum (Lambert et al. 2005) and activates the cytokine TNF-$\alpha$. This positive feedback loop leads to even greater local Ang1-8 production.

3) At the same time, Ang1-8/ATR1 enhances the inflammation:

3a) Ang1-8/ATR1 activates the KKS, with different effects on the two receptors BKR1 and BKR2. Through IL1 and TNF-$\alpha$, ATR1 upregulates the BKR1 receptor of DABK, whose level may increase through the decrease of ACE2, activating the DABK/BKR1 axis. Both ATR1 and BKR1 (Parekh et al. 2020) upregulate ADAM17-TACE, producing another positive feedback loop. At the same time, the other bradykinin receptor BKR2 is sensitized by ATR1, with which it forms a dimer (Quitterer and AbdAlla 2014). However, increased ACE downregulates the ligand BKR2, so the effect on BKR2 signalling is unclear. The net effect of the activation of the KKS is vasodilation, increased capillary leakage and recruitment of neutrophils.

3b) Ang1-8/ATR1 upregulates the Vascular Endothelial Growth Factor (VEGF) through TGF-beta and Angiopoietin-2, thus amplifying capillary leakage. At the same time, VEGF may be also activated by BKR2.

3c) Ang1-8/ATR1 and DABK/BKR1 recruit macrophages and neutrophils to the infection.

4) At some point, ACE2 is upregulated and it contributes to reversing the inflammation in multiple ways: ACE2 degrades Ang1-8 and Ang1-10, downregulating the Ang1-8/ATR1 axis, produces Ang1-7, upregulating the anti-inflammatory Ang1-7/Mas, and degrades DABK, diminishing vascular leakage. Downregulation of ATR1 reduces the VEGF, de-sensitives the BKR2 receptor, reduces the BKR1 receptors, reduces ADAM17-TACE and, through it, reduces TNF-$\alpha$.

Consistent with this view, ACE2 is upregulated during inflammation (Hanafy et al. 2011), in part through interferon I stimulation (Ziegler et al. 2020). ACE2 upregulation is also observed in a recent analysis of mRNA expression of bronchoalveolar lavage fluid cells (Garvin et al. 2020) and lung cells (Wu et al. 2020) of Covid-19 patients. However, the ACE2/ACE ratio is decreased in most chronic inflammatory diseases, and it may contribute to these pathologies (Li et al. 2020; Pagliaro and Penna 2020). As discussed below, we hypothesize that epigenetic mechanisms have a role in the discrepancy between short term activation and chronic reduction of ACE2.

**Severe Covid-19 as a failure of ACE2 to revert inflammation**

As discussed above, ACE2 plays a key role in reverting the inflammatory process in the context of the RAS and the KKS, and its downregulation favour endothelial damage, capillary leakage and angioedema, as shown by Imai et al. (2005) and Kuba et al. (2005) among others.

It has been shown that SARS-CoV (Haga et al. 2008) and SARS-CoV-2 degrade ACE2. We hypothesize that the degradation of ACE2, especially in presence of low initial levels, or the failure to upregulate ACE2 possibly due to low levels of Interferon (Bastard et al. 2020) may prevent the termination of the inflammatory process and perpetuate the propagation of inflammation through the affected organs, leading to organ damage and severe manifestations that can be life threatening.

In a transcriptomic study of lung tissue obtained from patients who died of Covid-19 in China it was found that the viral load was low in all samples. According to the authors, this "suggests that the patient deaths may be related to the host response rather than an active fulminant infection."

A retrospective study of almost 48,000 Covid-19 patients discharged from English hospitals (and, accordingly, negative to the virus in their large majority) found rates of hospital readmission and death 3.5 and 7.7 times greater, respectively, than in age- and comorbidities-matched controls, and higher rates of respiratory, diabetes and cardiovascular events, evidencing elevated rates of multi-organ dysfunction in individuals discharged from hospital (Ayoubkhani et al. 2021). Relative rates were higher for persons below 70 years than for older ones, and for ethnic minorities than for the White population. These observations indicate that many severe consequences of Covid-19 arise from the immune-inflammatory response rather than being a direct consequence of SARS-COV-2 infection.

**Variation of ACE2 across age**

Several studies based on RNA-seq experiments found that the mRNA of ACE2 is absent in the early foetus and it is lower in children than adults (Bunyavanich, Do and Vicencio 2020; Muus et al. 2020; Sharif-Askari et al. 2020; Inde et al. 2020). A similar observation applies to the ACE2 protein in serum (Pavel et al. 2020). Therefore, it was proposed that the low expression of the virus receptor ACE2 hinders the propagation of SARS-CoV-2 in children organs. However, several studies, including also some of those cited above, indicate that the expression of ACE2 with age is not monotonic, since ACE2 expression decreases at advanced age in several organs after reaching a maximum. At the mRNA level, decrease of ACE2 mRNA at advanced age was observed in mouse lung samples (Booeshaghi and Pachter 2020) and in an analysis of several human tissues collected in the GTeX database (Chen et al. 2020). Single cell data of human respiratory cells reported in Figure 3g of Muus et al. (2020) are also consistent with decrease or stationarity of ACE2 mRNA after a maximum reached at 10-25 years in multiciliated cells and 25-40 years in AT2 cells. A recent preprint (Inde et al. 2020) also found a maximum of ACE2 mRNA in mouse lungs at few days after birth, evidenced a minimum at approximately 10 days and an increase at least until 9-11 months, which in mice is middle age. For human data, the same preprint found a maximum of ACE2 mRNA short after birth in the hearth and at about 10 years in the testis (Fig. 6C and 6D of Inde et al.). Inde et al. also present data of ACE2 mRNA in human lungs in their Fig. 3G, which does not show the existence of a maximum but cannot rule it out due to the sparsity of data at advanced age.

Concerning ACE2 protein in cells, which is the relevant molecular species for virus propagation, western blot analysis of rat lungs (Xie et al. 2006) and mouse aorta (Yoon et al. 2016) indicate that ACE2 membrane protein levels decrease with age in adult rodents. These studies did not include juvenile animals and examined three age classes: 2 (3 for rats), 12 and 24 months, corresponding to young adults, middle age and old. In mouse thoracic aorta, the whole anti-inflammatory arm of the RAS to which ACE2 belongs decreases with age, while the pro-inflammatory arm of the RAS increase with age (Yoon et al. 2016). In humans, a study found that children below 10 years have on the average higher ACE2 protein in AT2 lung cells than adults, although these values vary considerably from cell to cell and from individual to individual (Ortiz-Bezara et al. 2020). However, this comparison depended on the discretization of ACE2 expression with an arbitrary threshold of more than 1% of positive cells, and it included adults with asthma that downregulates ACE2, whose exclusion would have raised the p-value. Moreover, other cell types did not confirm the same trend as AT2 lung cells. Zhang, Guo et al. (2020) found that ACE2 positive cells in lung biopsy samples from 26 children and 24 adults were significantly decreased in patients older than 50 with respect to children. This reduction was observed in bronchial cells but it was not significant in pulmonary alveolar cells, and it was also observed at the mRNA level. The study by Inde et al. (2020) also found extreme intra- and inter-individual heterogeneity, but it reached the opposite conclusion that ACE2 protein in human lung epithelial cells increases with age; however, also this conclusion was reached measuring the percentage of cells that express ACE2 above a threshold. Inde et al. also plot ACE2 intensities versus the donor age in their figures 1C and 1G, but these figures do not suggest a clear increase of ACE2 with age. Furthermore, they did not observe increase of ACE2 with age in AT2 cells, which are the cell type with maximum ACE2 expression in the lungs (Li et al. 2020).

We conclude that most existing data are consistent with a decrease across age of ACE2 mRNA and protein levels in rodents and several human cell types. Accordingly, we plot in Fig.3A a model in which ACE2 expression starts in the last stage of foetal life, reaches a maximum at young age and then decreases at old age, with data inspired on the data of Muus et al. in multiciliated cells. The position of the maximum depends on the cell type and it is influenced by the strong inter- and intra-individual variability evidenced by several studies (Ortiz-Bezara et al. 2020; Inde et al. 2020).

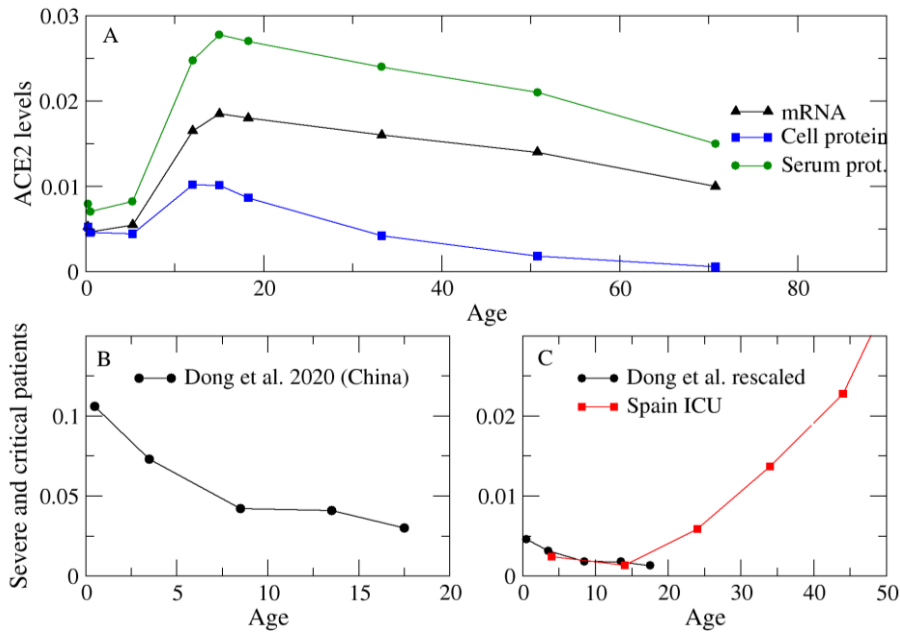

Fig.3. A: Proposed behavior of ACE2 with age at the three levels of mRNA, protein in the cellular membrane and protein in serum. B: The severity of pediatric cases of Covi-19 in China decreases with age (Dong et al. 2020); C: The same data, rescaled to take into account undetected cases, and put in the context of severe Covid-19 cases of children and adults from Fig.1.

For understanding the relation between ACE2 mRNA and protein expression it is necessary to consider the expression of the metalloprotease ADAM17/TACE. This enzyme sheds ACE2 from the cell membrane to the serum (Lambert et al. 2005; Xu et al. 2017), from which it is rapidly lost through urine. The expression of ADAM17 increases with age (Dou et al. 2017; Liu et al. 2019), consistent with the increase with age of Ang1-8 and its receptor ATR1 (Yoon et al. 2016) that upregulates ADAM17 (Deshotels et al. 2014) together with the bradykinin receptor BKR1 (Parekh et al. 2020).

As a result of mRNA expression and shedding through ADAM17, the stationary concentration of ACE2 cell protein is proportional to the ratio ACE2mRNA/ADAM17. We plot this function in Fig.3A, where we assume that ADAM17 decrease is inversely proportional to the square root of age. One can see that the maximum of ACE2 cell protein is reached at earlier age than the maximum of ACE2 mRNA, as the consequence of the increase of ADAM17 with age. ACE2 cell protein decreases in rats and mice from 3 to 12 months (Xie et al. 2006; Yoon et al. 2016) while ACE2 mRNA increases in mice of the same age (Inde et al. 2020). Although this comparison involves different species (rats versus mice) or different cell types (aorta versus lungs), it is consistent with the above prediction. ACE2 protein in the serum is proportional to the product of the cell protein times the shedding rate, i.e. it is proportional to ACE2 mRNA averaged over the different cell types. We depict in Fig.3A the proposed behaviour of ACE2 versus age at the levels of mRNA, cell protein and serum protein.

**Covid-19 severity across age presents a minimum**

A paediatric study in China (Dong et al. 2020) observed that the severity of Covid-19 in children (not including the MIS-C syndrome) decreases with age: the proportion of severe and critical paediatric cases was 10.6%, 7.3%, 4.2%, 4.1%, and 3.0% for the age groups <1, 1-5, 6-10, 11-15 and 16-18 years, respectively (Fig.3B). This decrease of severity from young children to teen agers is consistent with the drop in the hospitalization rate in Spain from the age class 0-10 to 10-20 that one can see in Fig.1. Covid-19 severity increase with age in adults (Fig 1), therefore severity versus age has a minimum (Fig.3C). Suggestively, the shape of this curve appears inversely related to the ACE2 protein level across age (Fig.3A), although the position of the maximum of ACE2 depends on the cell type. Another recent paper analysed similar data but proposed a direct instead of inverse relationship between Covid-19 severity and ACE2 expression (Inde et al. 2020). However, most existing data suggest that ACE2 expression decreases at high age, although this point is still debated, and the increase of severity with age is a very clear property of Covid-19.

**Comorbidities support a negative relation between ACE2 and Covid-19 severity**

As discussed in more detail in a preprint (Bastolla 2020), this inverse relationship between ACE2 levels and Covid-19 severity might rationalize not only the influence of age and sex but also most of the other known risk factors of Covid-19 examined in the OpenSAFELY study (Williamson et al. 2020), including diabetes, hypertension, obesity, vitamin D deficit (that may explain at least part of the risk factors connected with ethnicity) and perhaps even the curious protecting effect of smoke that increases ACE2 expression (Sharif-Askari et al. 2020).

Besides being low in most chronic inflammatory diseases (Li et a. 2020; Pagliaro and Penna 2020), the ACE2/ACE cell protein was found to be lower than control in broncho alveolar lavage (BAL) samples from 14 pediatric patients with acute respiratory distress syndrome (ARDS) (Wösten-van Asperen et al. 2013) and in mice with ARDS, in which the ACE2/ACE ratio decreased with age (Schouten et al. 2016). The decrease of ACE2/ACE with age was not observed in human patients with ARDS (Schouten et al. 2019), but this observation was based on only one measure per patient that might have been obscured by the complex time course of the disease.

**Effect of receptor level on viral propagation**

The theory according to which lower ACE2 level may protect from severe disease is based on the premise that lower receptor level hinders the propagation of the virus. But is this assumption warranted? Mathematical models of virus propagation, tested with experiments with bacteriophages, suggest that there is an optimal receptor density at which the virus propagates fastest, with viral velocity declining both for lower and for higher receptor density (Ortega-Cejas et al. 2004; see Fig. 4); and a mathematical model of the dependence of Covid-19 mortality on receptor density fitted to both Covid-19 and SARS 2003 data of lethality across age and sex suggests that both viruses propagates in a regime in which increasing receptor density slows down their propagation (Bastolla 2020). Thus, the inverse relation between ACE2 expression and Covid-19 severity is not implausible.

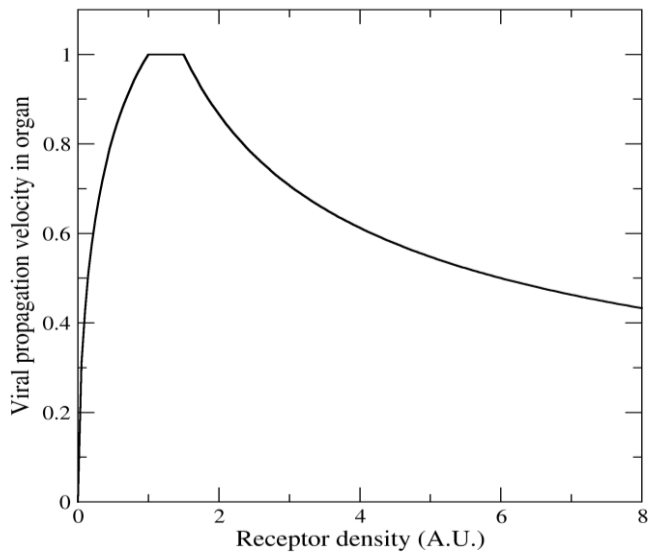

Fig.4 Mathematical model of the velocity of viral propagation in an organ as a function of the receptor density (adapted from the formulas presented in Ortega-Cejas et al. 2004 tested with experiments on bacteriophage propagation).

**The ATR2 receptor**

The proposed inverse relationship between ACE2 expression and Covid-19 severity has an exception in children, who experience mild Covid-19 but express less ACE2 than adults. However, children express more than adults the Ang1-8 receptor ATR2, which competes with ATR1 and counteracts its inflammatory effect (Millan et al. 1991; Sechi et al. 1992; Shanmugam et al. 1996; Kaschina and Unger 2003; Kaschina, Namsolleck and Unger 2017; Yoon et al. 2006). The decrease of ATR2 with age has been recently disputed (Gao et al. 2012), but most works support it, and they also concluded that the receptor binding activity of ATR2, which is most relevant in this context, decreases with age. Both ATR1 and ATR2 are GPCR, and they can form complexes between themselves and with other receptors, including BKR2, underscoring complex regulatory interactions and downregulation of ATR1 by ATR2 (Inuzuka et al. 2016).

We propose that the high level of the ATR2 receptor in children may compensate the low ACE2 level for responding to increased levels of Ang1-8, in particular by defusing the positive loops that amplify Ang1-8. Ang1-8 bound to the pro-inflammatory receptor ATR1 upregulates ACE and downregulates ACE2 at the mRNA level (Koka et al. 2008) and upregulates ADAM17 that sheds ACE2 from cells (Deshotels et al. 2014), thereby increasing its production rate by ACE and reducing its degradation by ACE2. These regulatory loops require activation of ATR1, and they are hindered by high expression of the alternative receptor ATR2. As the level of ATR2 decreases with age, the positive feedback loops become stronger and may contribute to the increase of Ang1-8 and in general the pro-inflammatory arm of the RAS across age (Yoon et al. 2016).

**The role of interferons**

It has been recently shown that type-I interferons (IFN-I) upregulate ACE2 expression (Ziegler et al. 2020). Interestingly, IFN-β are used as anti-inflammatory therapy in multiple sclerosis (Kieseier 2011), supporting their anti-inflammatory role. We conjecture that the upregulation of ACE2 by IFN-I may contribute to stop the inflammatory cascade. This conjecture is consistent with the recent finding that auto-antibodies against IFN-α are associated with more severe forms of Covid-19 (Bastard et al. 2020) and that impaired type-I IFN response (Hadjadj et al- 2020) and genetic variants that hinder interferon activation (Zhang et al. 2020) are associated with severe Covid cases. Moreover, mutations of the IFN-I receptor have been found to be risk factors of severe Covid-19 through genome wide association studies (GWAS) analysis (Pairo-Castineira et al. 2020). Finally, SARS-COV-2 infection is characterized by low levels of IFN-I (Blanco-Melo et al. 2020), partly because the virus dysregulates IFN-I (Konno et al. 2020). Therefore, we suggest that IFN downregulation cooperates with ACE2 dysregulation by SARS_COV-2 in preventing the arrest of the inflammatory process in severe Covid-19 cases. However, interferon therapy did not provide any reduction of mortality or hospitalization in the Solidarity clinical trial (WHO Solidarity Trial Consortium 2020).

**Epigenetic enhancement of the inflammatory process**

An increasing amount of evidence indicates that inflammation related genes in general, and ACE2 in particular, are subject to epigenetic control through chromatin modifications. Transcriptional memory provides faster and enhanced transcription upon repeated stimulations, and it is a common hallmark of interferon-stimulated genes, mediated by histone variants such as H3.3 and histone modifications that favour transcription (Kamada et al. 2018) and by de-methylation of CpG islands at active promoters, associated with sustained stimulation by the proinflammatory cytokine TNF-α (Zhao et al. 2020).

The ACE2 gene has been shown to be subject to transcriptional memory through CpG island demethylation that increases through age (Corley and Ndhlovu, 2020), and ACE2 transcription in samples of patients with comorbidities associated with severe COVID-19 was correlated with the expression of genes related to histone modifications, such as HAT1 (type B histone acetyltransferase involved in the acetylation of newly synthesized histones), HDAC2 (histone deacetylase involved in transcriptional repression) and KDM5B (histone demethylase involved in transcriptional repression) (Pinto et al. 2020). Genome-wide CRISPR screens identified several chromatin modifiers that facilitate SARS-COV-2 infection (Wei et al. 2020), likely by upregulating ACE2: the gene HMGB1 encoding a DNA binding protein that regulates chromatin, the SWI/SNF nucleosome remodeling complex that regulates chromatin accessibility and gene expression, the histone demethylase KDM6A, the histone methyltransferase KMT2D and the lysyl hydroxylase JMJD6, while the histone H3.3 chaperone complex HUCA hindered viral infection, suggesting an anti-viral role for deposition of the histone variant H3.3.

The transcriptional memory of the ACE2 gene has been interpreted by Pruimboom et al. (2020) as evidence of a positive correlation between ACE2 expression and Covid-19 severity. However, it is important to note that the transcriptional memory also acts on pro-inflammatory genes (Kamada et al. 2018; Zhao et al. 2020). The reported decrease of the ACE2/ACE ratio in chronic inflammatory diseases associated with severe Covid-19 (Li et al. 2020; Pagliaro and Penna 2020) suggests that the pro-inflammatory side prevails in this balance, with the final result to decrease the ACE2/ACE ratio through the transcriptional downregulation of ACE2 (Koka et al. 2008), which is also consistent with the decrease of the ACE2/ACE ratio observed in ARDS (Wösten-van Asperen et al. 2013; Schouten et al. 2016) and with the theory of inflammaging, which postulate that the increase through age of inflammatory processes contributes to the phenomenology of aging and senescence (Franceschi et al. 2007; Minciullo et al. 2016).

Future work should address the complex regulation of ACE2 and other inflammatory and anti-inflammatory genes from a system perspective, considering multiple level of regulation (epigenetic, transcriptional, post-translational).

**Therapeutic perspectives**

Our hypothesis supports several candidate drugs that might alleviate severe complications of CoViD-19 and even its post infection consequences (MIS-C and long Covid-19) by inhibiting the processes that prevent the reversion of the inflammatory state. Luckily for several of these drugs clinical trials are already on-going.

1) human recombinant ACE2. It can have a double protective effect. On one hand, it can alleviate the inflammatory process, and in the other one it can sequester the virus and difficult its entry in the cells (Batlle et al. 2020). A recent case report presents very promising results, and clinical trials are ongoing (Zoufaly et al. 2020).

2) Angiotensin receptor blockers (ARB) that block the ATR1 receptor, downregulating the inflammatory process and preventing the positive feedback loop of Ang1-8 through upregulation of ACE and downregulation of ACE2. They are often discussed together with ACE inhibitors (ACE-I) that reduce the formation of Ang1-8. However, ACE has anti-inflammatory effect on the KKS, where it downregulates bradykinin, so that one of the known adverse effects of ACE-I is the upregulation of KKS eventually leading to edema. Moreover, Ang1-8 can be also formed through other peptidases.

The possible therapeutic role of ARB and ACE-I was suggested by several groups already at the time of SARS 2003 (Sun et al. 2020; Gurwitz 2020; Verdecchia et al.

2020; Ciaglia et al. 2020; Offringa et al. 2020; Annweiler et al. 2020). Although there was the concern that, by upregulating ACE2, ARB and ACE-I might favour the virus, meta-analysis of several clinical studies suggested that they are associated with lower mortality of COVID-19 in patients with hypertension (odds ratio, 0.57 [95% CI, 0.38–0.84], Guo, Zhu and Hong 2020). The clinical trials NCT04312009 and NCT04311177 are ongoing to test possible protective effects, although caution is necessary if blood pressure becomes too low, which is a frequent consequence of severe Covid-19 (Michard and Vieillard-Baron 2020, Lala et al. 2020), probably through the activation of the bradykinin axis.

3) Drugs that enhance the ATR2 receptor. Their effect may be in principle similar to ARB.

4) The product of ACE2, Ang1-7, which exerts anti-inflammatory and anti-fibrotic effects (Chappell MC and Zayadneh 2017) through the receptor Mas. However, its mechanism of action is still not fully known, it is rapidly degraded by ACE, and it has a moderate affinity for ATR1, thus it might produce effects contrary to the intentions;

5) IFN-I that upregulate ACE2. Although interferon may promote inflammation, IFN-β (type I) is used as anti-inflammatory drug to treat multiple sclerosis (Kieseier 2011). However, the Solidarity clinical trial did not observe any reduction of Covid-19 mortality upon IFN-I treatment (WHO Solidarity Trial Consortium 2020).

6) Other drugs that upregulate ACE2

7) Inhibitors of IL-1 and TNFalpha, which are successfully used to treat KD and MIS-C, might also be useful to treat Covid-19 for preventing excessive activation of the KKS.

8) Vitamin D, which exerts a protective role by downregulating ATR1 and ADAM17, and in this way limits the inflammatory process and protects ACE2 from excessive shedding. A retrospective study of almost 5 million people found an association between vitamin D deficiency and Covid-19 severity (Israel et al. 2020), and a clinical study with 76 patients found that a high dose of Calcifediol, a metabolite of vitamin D, significantly reduced the ICU treatment for hospitalized Covid-19 patients (Entrenas-Castillo et al. 2020).

Of course, experimental verifications of the above hypothesis are necessary, and clinical trials are urgently needed to test these therapeutic indications.

**Conclusions**

As reviewed above by us and elsewhere by many other authors, the peptidase ACE2, besides being the cellular receptor of SARS-COV-2 and other coronavirus, exerts a central regulatory role in inflammatory processes by downregulating the main pro-inflammatory peptides of the Renin-angiotensin system (RAS) and the Kinin kallikrein system (KKS) (Fig.2). These two highly integrated signalling systems, whose receptors act synergistically and strengthen each other, activate the inflammatory response, activate cytokines starting from a central controller of inflammation such as TNF-$\alpha$, which is activate by the same protease TACE/ADAM17 that degrades ACE2, and recruit macrophages and neutrophils. We propose that the main role of ACE2 consists in terminating this inflammatory process. Failure to do so, because of ACE2 degradation or failure to activate it, may exacerbate the inflammatory response leading to organ damage, including edema, and it may be responsible of the most severe consequences of Covid-19 even after the infection has been resolved, as it happens in the Multi-inflammatory syndrome of children (MIS-C).

Because of the central role of ACE2 both for SARS-COV-2 propagation and for controlling the inflammatory process, and because of the very marked increase of Covid-19 with age (Fig.1), a strong effort has been dedicated to elucidate how ACE2 expression change with age. However, these works have produced conflicting results, that have been interpreted either as support of the theory that the receptor expression favours SARS-COV-2 infection or as support of the opposite theory that high ACE2 expression favours milder disease because of its anti-inflammatory role.

We have shown that these contrasting results on ACE2 expression across age can be reconciled going beyond the dichotomy between adults and children and recognizing that ACE2 expression does not change monotonically with age but it starts in late foetal life, reaches a maximum at a young age that depends on the cell type and exhibits strong inter-individual and inter-cellular variation, and decreases at advanced age. At the protein level, this decrease starts before and it is more accentuated than at the mRNA level, due to the increase with age of the activity of the protease TACE/ADAM17 that sheds ACE2 from the cell membrane to the serum. This proposed behaviour, represented in Fig.3A, is consistent with essentially all existing data on the ACE2 expression across age in rodents and humans, except some of the figures of the preprint by Inde et al. (2020).

The complexity and the intricacies of ACE2 regulation will require future approaches that integrate multiple regulatory levels, from chromatin remodelling that plays a key role in transcriptional memory to regulation through transcription factors and post-translational regulation through ADAM17 and other proteases that degrade ACE2. This is an important subject, since the decrease of ACE2 with age, and the simultaneous increase of the pro-inflammatory components of the RAS may play an important role in the theory of inflammaging, which postulate that the increase through age of inflammatory processes contributes to the phenomenology of aging and senescence (Franceschi et al. 2007; Minciullo et al. 2016).

The proposed decrease of ACE2 across age, and the observed decrease of the ACE2/ACE ratio in chronic inflammatory diseases often associated with severe Covid-19 (Li et al.2020; Pagliaro and Penna 2020) support a negative relationship between Covid-19 severity and ACE2 expression (Fig.3B). This is not as paradoxical as it seems, since a mathematical model of virus propagation across cells predicts that an increase of the receptor level does not necessarily accelerate viral propagation and it may even slow it down (Ortega-Cejas et al. 2004; see Fig.4 for illustration).

The exception to this proposed negative relationship is constituted by children, who express less ACE2 than adults and experience much milder disease. However, in children the RAS is much less dis-balanced towards the pro-inflammatory side than in adults. In particular children express more than adults the alternative angiotensin receptor ATR2, that exerts an anti-inflammatory role by downregulating the pro-inflammatory receptor ATR1, competing with it for angiotensin II binding, and reducing the positive feedbacks that downregulate ACE2 and enhance its degradation by ADAM17.

Of course, experimental work is needed for supporting the hypothesis presented in this review, but we hope that they may help rationalizing apparently conflicting observations.

## Acknowledgements

We gratefully acknowledge financial support from the Spanish Research Council (CSIC) under the grant CSIC-COV19-108. We thank Dr. Juan Carlos Alonso for interesting discussions.